\documentclass[]{article}
\usepackage{multicol}
\usepackage{apj}
\def\lesssim{\mathrel{\hbox{\rlap{\hbox{\lower4pt\hbox{$\sim$}}}\hbox{$<$}}}}
\def\gtrsim{\mathrel{\hbox{\rlap{\hbox{\lower4pt\hbox{$\sim$}}}\hbox{$>$}}}}
\newcommand{\mincir}{\raise -2.truept\hbox{\rlap{\hbox{$\sim$}}\raise5.truept
\hbox{$<$}\ }}
\newcommand{\magcir}{\raise -2.truept\hbox{\rlap{\hbox{$\sim$}}\raise5.truept
\hbox{$>$}\ }}
\newcommand{\siml}{\raise -2.truept\hbox{\rlap{\hbox{$\sim$}}\raise5.truept
\hbox{$<$}\ }}
\newcommand{\simg}{\raise -2.truept\hbox{\rlap{\hbox{$\sim$}}\raise5.truept
\hbox{$>$}\ }}
\newcommand{\be}{\begin{equation}}
\newcommand{\ee}{\end{equation}}
\newcommand{\ba}{\begin{eqnarray}}
\newcommand{\ea}{\end{eqnarray}}

\newcommand{\hm}{\,h^{-1}{\rm Mpc}}
\newcommand{\vel}{\,{\rm km\,s^{-1}}}
\newcommand{\fl}{\,{\rm erg\,s^{-1}cm^{-2}}}
\newcommand{\lum}{\,{\rm erg\,s^{-1}}}

\begin{document}

%%%%%%%%%%%% per formato preprint
\vspace{15mm}                                                                   
\begin{center}
\uppercase{Velocity Dispersions of CNOC Clusters\\ 
and the Evolution of the Cluster Abundance}\\
\vspace*{1.5ex} 
\sc{
Stefano Borgani{
\footnote{\label{1} INFN, Sezione di Perugia, c/o Dipartimento di
Fisica dell'Universit\`a, via A. Pascoli, I-06123 Perugia,
Italy}
\footnote{\label{2} INFN, Sezione di Trieste, c/o Dipartimento di
Astronomia, Universit\`a degli Studi di Trieste, via Tiepolo 11,
I-34131 Trieste, Italy; borgani@ts.astro.it.}}, 
Marisa Girardi{
\footnote{\label{3} Dipartimento di Astronomia, Universit\`a degli Studi di
Trieste, via Tiepolo 11, I-34131 Trieste, Italy; girardi@ts.astro.it.}},
Ray G. Carlberg{
\footnote{\label{4} Department of Astronomy, University of Toronto,
Toronto ON M5S 3H8, Canada; carlberg@astro.utoronto.ca.,
hyee@astro.utoronto.ca.} 
\footnote{\label{5} Visiting Astronomer, Canada-France-Hawaii
Telescope, which is operated by the National Research Council of
Canada, Le Centre National de Recherche Scientifique, and the
University of Hawaii}}
\\Howard K.C. Yee{$~^{\ref{4}}$ $^{\ref{5}}$}
and Erica Ellingson{$~^{5}$
\footnote{\label{6} Center for Astrophysics and Space Astronomy,
University of Colorado, Boulder, CO 80309, U.S.A.;
e.elling@casa.colorado.edu.}} 
}\\
\vspace*{1.ex} 
\end{center}
\vspace*{-6pt}

\begin{abstract}
We present the results of the analysis of the internal velocity
dispersions, $\sigma_v$, for the sample of 16 distant galaxy clusters
($0.17\mincir z\mincir 0.55$) provided by the Canadian Network for
Observational Cosmology (CNOC). Different $\sigma_v$ estimates are
provided, all based on an interlopers removal algorithm, which is
different from that originally applied by Carlberg et al. (1996). We
find that all such methods provide $\sigma_v$ estimates which are
consistent within $< 10\%$ among themselves and with the original
estimates provided by the CNOC collaboration. This result points in
favor of a substantial robustness of currently applied methods for
optical studies of the internal cluster dynamics. The resulting
distribution of velocity dispersions is used to trace the redshift
evolution of the cluster abundance with the aim of constraining the
matter density parameter, $\Omega_m$.  We find that constrains on
$\Omega_m$ are very sensitive to the adopted value of $\tilde
\sigma_8=\sigma_8\Omega_m^\alpha$ ($\alpha\simeq0.4$--0.5), as
constrained by the local cluster abundance. We find that, as
$\tilde\sigma_8$ varies from 0.5 to 0.6, the best fitting density
parameter varies in the range $0.3\mincir \Omega_m\mincir 1.0$. A
further source of uncertainty in constraining $\Omega_m$ is due to
uncertainties in the correction for the $\sigma_v$--incompleteness of
the CNOC sample. This calls for the need of better understanding the
constraints from the local cluster abundance and increasing the
statistics of distant clusters in order to suppress the systematics
related to the sample completeness criteria.

%\end{abstract}
%\keywords{ galaxies: clusters: general -- galaxies: fundamental 
%parameters -- cosmology: observations}

%%%%%%%%%%%formato preprint
%
\vspace*{6pt}
\noindent
{\em Subject headings: }
Cosmology: galaxies - clusters - theory - large-scale
structure of the universe.
\end{abstract}

%%%%%%%%%%%% per formato preprint
\begin{multicols}{2} 
%%%%%%%%%%%%

\section{INTRODUCTION}
Galaxy clusters represent the virialization stage of exceptionally high
peaks of initial density perturbations on comoving scales of about
$10\hm$\footnote{Here $h$ is the Hubble constant in units of
100$\vel$ Mpc$^{-1}$.}. Therefore, their abundance is
sensitive to the amplitude of the fluctuation power spectrum on such
scales. Standard analytical methods based on the approach originally
devised by Press \& Schechter (1974, PS hereafter) show that the
number density of clusters of a given mass provides a robust
constraint on $\tilde \sigma_8=\sigma_8\Omega_m^\alpha$, where
$\sigma_8$ is the r.m.s. fluctuation amplitude within a sphere of
$8\hm$ radius, $\Omega_m$ is the matter density parameter and
$\alpha\simeq 0.4$--0.5, weakly dependent on $\Omega_m$, on the
presence of a cosmological constant term and on the shape of the
power--spectrum (e.g., White, Efstathiou \& Frenk 1993). 

However, while theoretical predictions provide the number density of
clusters as a function of their mass, observations give the cluster
abundance as a function of some observable quantity, like the $X$--ray
luminosity, the $X$--ray temperature or the velocity dispersion of
member galaxies, which are {\em a posteriori} connected to mass at
different degrees of reliability.  Despite the variety of methods and
of data sets employed to trace the local cluster abundance, all the
analyses consistently indicate that $\tilde\sigma_8\simeq 0.5$--0.6
(Eke et al. 1996; Kitayama \& Suto 1997; Oukbir, Bartlett \& Blanchard
1997; Girardi et al. 1998a; Pen 1998; Markevitch 1998; Borgani et
al. 1999, B99 hereafter; Viana \& Liddle 1999). Once a model is tuned
so as to reproduce the abundance of local clusters, its evolution
mainly depends on $\Omega_m$ (e.g., Oukbir \& Blanchard
1992). Therefore, having a statistical sample of high--redshift
clusters with reliable mass determinations would allow in principle to
break the degeneracy between $\Omega_m$ and $\sigma_8$.

The growing availability of high--redshift cluster samples selected in
the $X$--ray band led in the last few years to a flurry of activity
along this line. Using the redshift distribution of clusters from the
Einstein Medium Sensitivity Survey (EMSS, Gioia et al. 1990), Sadat,
Blanchard \& Oukbir (1998) found consistency with a critical--density
($\Omega_m=1$) Universe (cf. also Reichart et al. 1999). B99 
analyzed the ROSAT Deep Cluster Survey
(RDCS, Rosati et al. 1995, 1998) and pointed out that current
uncertainties in the evolution of the mass--luminosity relation
prevent one from obtaining strong conclusions on $\Omega_m$ even from a
flux--limited sample as deep as RDCS ($z\mincir 1$). 

In this respect, the possibility of measuring $X$--ray cluster
temperatures would circumvent this problem to a fair degree, 
since it is more directly connected to the mass
than the $X$--ray luminosity. Indeed, Eke et al. (1998) analyzed the
Henry (1997) sample of $X$--ray cluster temperatures, extending out to
$z= 0.33$. They concluded that the $X$--ray temperature function (XTF)
favor a low--density Universe, with $\Omega_m\simeq 0.4\pm 0.2$.
Quite remarkably, Viana \& Liddle (1999) analyzed the same data set
and claimed that, after accounting for systematics in both the data
set and in the theoretical PS framework, a critical density Universe
is still viable as far as the evolution of the XTF is concerned. 

It is clear that such ambiguities can in principle be eliminated by
resorting to cluster data at a substantially higher redshift, where
differences among different Friedmann geometries rapidly
increase. However, the price to be paid in this case is that a much
smaller number of clusters with highly accurate data is presently
available (e.g., Luppino \& Gioia 1995; Donahue et al. 1998; Rosati et
al. 1999).

An alternative way to trace the cluster mass function back in redshift
is offered by measurements of the velocity dispersions, $\sigma_v$, of
member galaxies.  This represents a well established technique, which
has been already extensively applied to samples of local clusters
(e.g., Zabludoff, Huchra, \& Geller 1990; Girardi et al. 1993; den
Hartog \& Katgert 1996; Fadda et al. 1996; Mazure et al. 1996; Borgani
et al. 1997, and references therein).

Systematic estimates of the velocity dispersions for a statistical
sample of distant clusters has been performed for the first time by
Carlberg et al. (1996, C96 hereafter). They analyzed the Canadian
Network for Observational Cosmology (CNOC) sample of clusters (Yee,
Ellingson \& Carlberg 1996), which includes 15 $X$--ray selected
clusters from EMSS and Abell 2390.  Carlberg et al. (1997b, C97
hereafter) used the CNOC sample to trace the evolution of the cluster
abundance and concluded that $\Omega_m$ values below unity are in
general to be preferred (see also Bahcall, Fan \& Cen 1997). However,
this result depends sensitively on ${(a)}$ the $z\simeq 0$
normalization from the number density of local clusters with velocity
dispersion above a given value, and ${(b)}$ on the procedure to
convert the distribution of cluster velocity dispersions into the
distribution of cluster masses. In particular, one of the questions
raised by C97 concerned the comparison between the different
algorithms applied to estimate $\sigma_v$ for local clusters and for
the distant CNOC clusters. If the estimator used for local clusters
provides $\sigma_v$ which are biased upwards, this would give an
overestimate of $\tilde\sigma_8$ and, therefore, of $\Omega_m$.

A reliable estimate of $\sigma_v$ faces several problems, such as the
presence of foreground and background interlopers, velocity anisotropy
in galaxy orbits, the presence of substructures and the limited amount
of data available.  Girardi et al. (1993) showed that different
methods presented in the literature to estimate $\sigma_v$ (Yahil \&
Vidal 1977; Zabludoff, Huchra \& Geller 1990; Beers, Flynn \& Gebhardt
1990) give similar results on well sampled clusters, while only the
robust estimator by Beers et al. (1990) seems to be efficient when
only $\sim 10$ galaxy redshifts per cluster are available. Besides
the estimate of $\sigma_v$ from redshifts of member galaxies, a
further critical issue concerns the identification of genuine cluster
members by removing foreground and background interlopers.

For instance, Girardi et al. (1993) rejected interlopers by following
a guess for the outer limits of the redshift range encompassed by each
cluster. Two different, more refined, methods for interloper removal
have been subsequently developed by Fadda et al. (1996, F96 hereafter;
see also Girardi et al. 1998b, G98 hereafter) and by den Hartog \&
Katgert (1996; see also Adami et al. 1996; Mazure et al. 1996), both
based on combining information on projected position and velocity of
each galaxy, although by using rather different procedures.  Quite
remarkably, such procedures of cluster member selection lead to
velocity dispersions which are in good agreement (cf. F96 and Adami et
al. 1998). Significant differences were found only for a small number
of clusters (8 out of 74) which were recognized by the procedure of
F96 as multiple structures in redshift space.

Despite their differences, a common feature of all such methods is
that they are based on the identification of individual interlopers to
be removed.  Quite differently, the method followed by C96 to estimate
$\sigma_v$ for CNOC clusters statistically subtract the mean density
of field galaxies from the redshift space of the cluster. The
resulting velocity dispersions, which are about 13\% lower than
without the background subtraction, have been shown by Mushotzky \&
Scharf (1997) to be in good agreement with those inferred from the
$X$--ray temperatures provided. Furthermore, Lewis et al. (1999) have
shown that virial masses for CNOC clusters are quite consistent with
those obtained from $X$--ray data analysis.

In this paper, we perform a complete re--analysis of the cluster
velocity dispersion of the CNOC sample by applying different
algorithms which are similar to that used by F96 for local
clusters. Such estimates of the velocity dispersions will be used to
estimate the evolution of the cluster abundance through the
redshift--dependence of the $\sigma_v$ distribution.

Therefore, the final goal of this paper will be to answer to the
following two questions.
{\em (a)} Does the explicit background subtraction method, applied by
C96 to estimate $\sigma_v$ for CNOC clusters,
provide consistent results to those from methods applied by F96 and
G98 to local cluster samples?
{\em (b)} Having a reliable determination of velocity dispersions for
a sample with the statistics and the redshift extension of CNOC,
are we able to provide robust constraints on $\Omega_m$
from the evolution of the cluster abundance?

The structure of the paper is as follows. In Section 2 we briefly
describe the CNOC sample. In Section 3 we introduce the different
methods to estimate $\sigma_v$. After showing the results of the
application to the CNOC clusters, we compare the results to those
provided by C96 (cf. also Carlberg et al. 1997a). In Section 4 we use
such results on $\sigma_v$ to place constraints on $\sigma_8$ and
$\Omega_m$. We draw our main conclusions in Section 5.

\section{THE CNOC SAMPLE}
\label{s:data}
A comprehensive description of the CNOC cluster sample is provided by
Yee, Ellingson \& Carlberg (1996). In the following we only describe
the main features which are relevant to our analysis, while we refer
to that paper for a more complete presentation.  The CNOC cluster
redshift survey includes all the 15 EMSS clusters (Gioia et al. 1990;
Henry et al. 1992) which have $L_X>4\times 10^{44}\lum$ ($h=0.5$ and
$q_0=0.5$) and $f>f_{lim}=4\times 10^{-13}\fl$ for $X$--ray
luminosities and fluxes in the $[0.3,3.5]$ keV energy band, redshift
in the range $[0.17,0.55]$ and $-15^\circ < \delta < +55^\circ$ for
the declination. The solid angle covered by EMSS in this declination
range is of 587 sq. deg. Also included in the original CNOC sample is
Abell 2390, which however does not belong to EMSS. Although it is
included in our $\sigma_v$ analysis, we will not use it for the
purpose of constraining cosmological models.

The CNOC surveys provided a fairly large number (50 to 200) of
galaxy redshifts per cluster over a region covering up to $6\hm$
projected on the sky. Galaxy magnitude selection criteria have been
chosen so as to minimize contamination due to foreground/background
objects. This makes the CNOC clusters very well suited to follow the
redshift evolution of the cluster internal dynamics. 

%%FIGURE 1%%%
%\end{multicols}
%\begin{figure}
\includegraphics{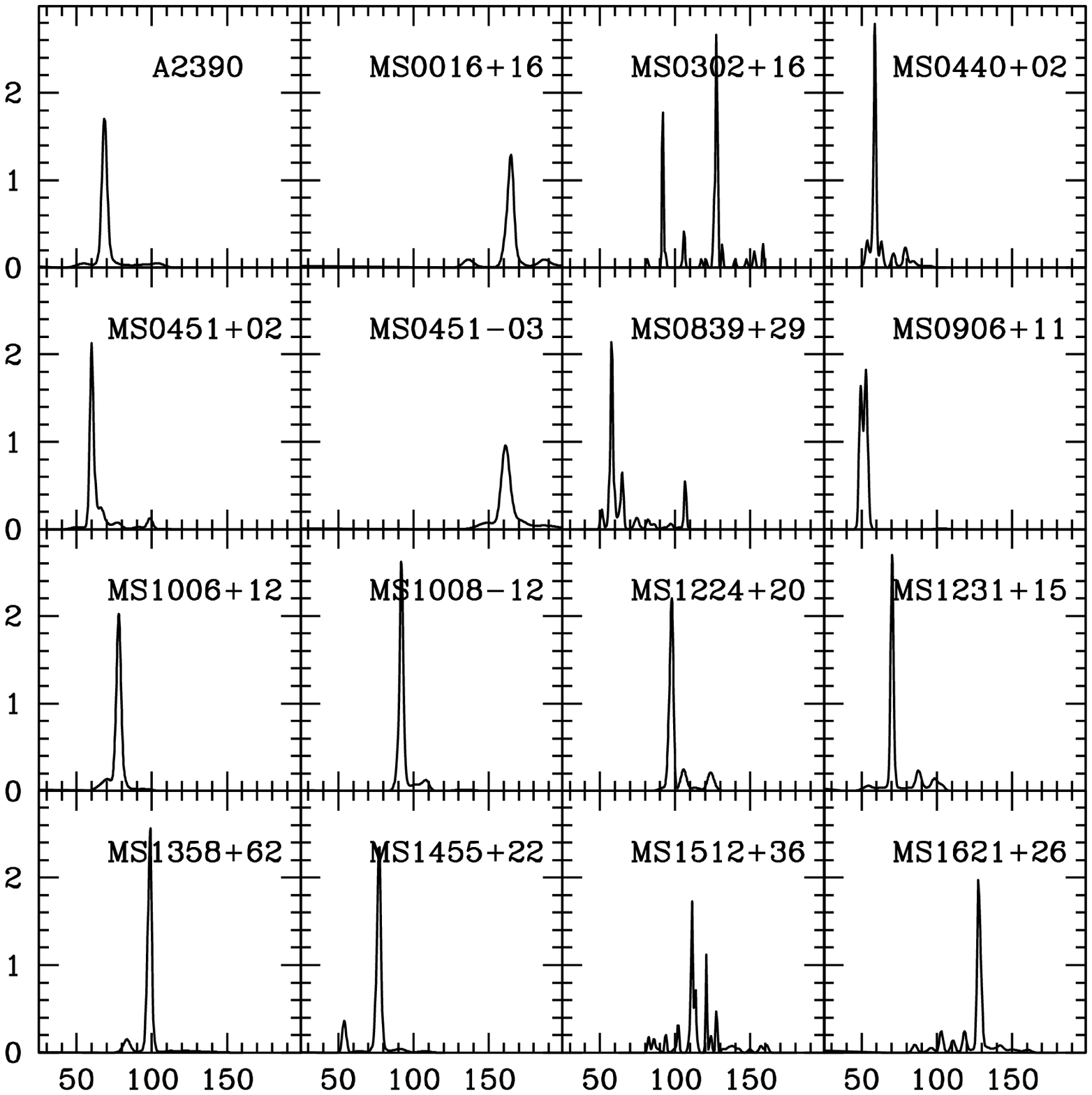}
$\ \ \ \ \ \ $\\
\vspace{9.4truecm}
$\ \ \ $\\
%\vspace{-20mm}
{\small\parindent=3.5mm {Fig.}~1.--- The velocity--space galaxy
density, as provided by the adaptive--kernel reconstruction
method. Units on the $x$--axis are velocities in $10^3 \vel$, while
the $y$ axis is in arbitrary units.  }

\vspace{5mm}
%\begin{multicols}{2}
%%%%%%%%%%%%%%

\section{Analysis of the CNOC sample}

In order to select member galaxies and compute velocity dispersions,
$\sigma_v$, we apply the same procedure adopted by F96 and G98 for
nearby clusters.  We remind that the other recent $\sigma_v$ analysis
for an extended sample of nearby clusters (Mazure et al. 1996) leads
to velocity dispersions which are in good agreement, although it
resorted to a different procedure for the selection of cluster members
(cf. F96 and Adami et al. 1998).

\subsection{Cluster Member Selection}

The member selection for the CNOC clusters is performed by considering
the whole sampled region around each cluster, with the exception of
MS1512+36, which we analyze only within 1$\hm$. This cluster appears
as very elongated and shows strong complexity in the two dimensional
galaxy distribution. Therefore, we prefer to avoid distant cluster
regions where this complexity makes difficult a dynamical analysis
(see below).  The initial list of candidate cluster members do not
include galaxies which are very far from the cluster redshift range and,
therefore, are clearly foreground/background objects.

The identification of cluster members proceeds in two steps. 

Firstly, we perform the cluster membership selection in velocity space
by using only redshift information. We use the adaptive kernel method
(Pisani 1993) to find the significant ($>99\%$ c.l.)  peaks in the the
velocity distribution. Only galaxies belonging to such peaks are
considered as candidate cluster members.  In Figure 1 we
plot the velocity--space galaxy density for all the CNOC clusters, as
provided by the adaptive--kernel reconstruction method.

From the above member selection analysis only the MS0906+11 cluster,
shows two significant peaks, which are separated by $> 3000 \vel$ in
velocity space (hereafter MS0906+11a and MS0906+11b).  According to
our procedure, these two peaks show a significant overlapping
($>40\%$) and they are not perfectly separable, thus leading to a
cluster with uncertain dynamics.  The amount of evidence of this binary
nature (e.g. C96; Carlberg et al. 1997a), as well as the strong
disagreement between $\sigma_v$ of the whole system ($1723 \vel$ from
the present analysis and $1893 \vel$ from C96) and the $X$--ray
temperature reported by Lewis et al. (1999; 8 keV, corresponding to
$\sim 1150 \vel$ under the hypothesis of energy equipartition between
gas and galaxies), indicates that this cluster should be better
considered as composed of two separated peaks.

Also the velocity distribution of MS1512+36 is very complex. The
cluster peak is surrounded by secondary peaks, which have a smaller
density and are separated from the main peak. This complexity further
motivates our choice to estimate $\sigma_v$ for this cluster only
within $1\hm$.

All those galaxies which pass through the first velocity--space
selection are analyzed in the second step, which uses the combination
of position and velocity information.  This procedure, which is based
on the ``shifting gapper'' algorithm, proceeds in two iterative steps.
Firstly, galaxies are assigned to radial bins of $0.4\hm$ width or
larger, in order to contain at least 15 objects. Secondly, those
galaxies that, within each bin, are separated in cluster rest--frame
velocity by $\ge 1000\vel$ from the main body of the velocity
distribution are identified as interlopers.  These two steps are
iterated until the number of cluster members does not change
anymore. Figure 2 shows the plots of rest--frame velocity
versus projected cluster--centric distance for those clusters where
the ``shifting gapper'' found interlopers (indicated with the open
circles).  We note that the MS1358+62 cluster contains a close system,
which corresponds to a southern group identified in velocity space and
already noted by C96 (the vertical line mark the separation between
these two systems). In order to exclude this secondary structure,
hereafter we analyze the MS1358+62 cluster only within $1.2 \hm$.

%%FIGURE 2%%%
%\end{multicols}
%\begin{figure}
\includegraphics{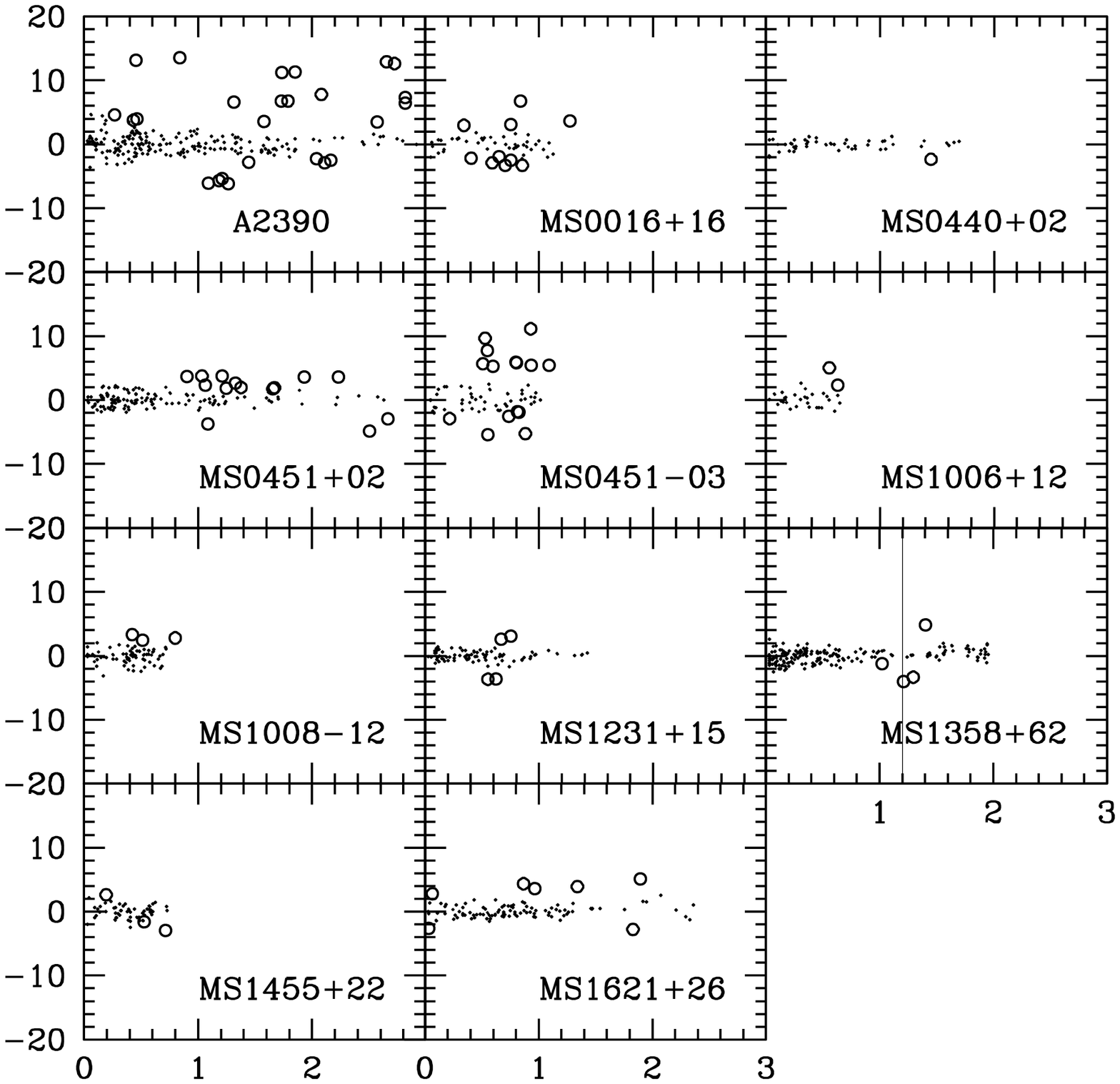}
$\ \ \ \ \ \ $\\
\vspace{9.4truecm}
$\ \ \ $\\
%\vspace{-20mm}
{\small\parindent=3.5mm {Fig.}~2.--- 
Rest--frame velocity versus projected cluster--centric
distance for each cluster where the ``shifting gapper'' succeeded in
rejecting at least one galaxy. Distances on the $x$--axis are in units
of $\hm$ and the rest-frame velocities on the $y$--axis are in
units of $10^3\vel$. Interlopers are indicated with the
open circles. As for the MS1358+62 cluster, the thin vertical line
marks the separation between the main cluster and a nearby subgroup
(see text).}

\vspace{5mm}
%\begin{multicols}{2}
%%%%%%%%%%%%%%

Finally, we estimate the corrections for velocity gradients, which are
however always quite small (e.g. Girardi et al. 1996). We verify that
only A2390 has significant (at $> 99\%$ c.l.) velocity gradients. For
this cluster we apply a correction by subtracting the velocity
gradient from each galaxy velocity and renormalizing the velocities so
as to have their mean velocity unchanged. This correction results in a
decrease of $\sigma_v$ (see below) by $58 \vel$.

\subsection{The velocity dispersion estimation}

We compute robust estimates of (line--of--sight) velocity dispersions,
$\sigma_{rob}$, by using the ROSTAT routines by Beers, Flynn, \&
Gebhardt (1990), after applying the relativistic correction
and the usual correction for velocity errors (Danese, De Zotti, \& di
Tullio 1980). We use the bi--weight scale estimator when more than
fifteen member galaxies are available, while the gapper estimator is
applied otherwise (see also Girardi et al. 1993), 

Several authors (F96; den Hartog \& Katgert 1996; Girardi et
al. 1996) have shown that velocity dispersion profiles for individual
clusters are often a strong, either increasing or decreasing,
function of the radius in the central cluster regions. This behavior
may be due both to velocity anisotropies and to the dark matter
distribution, although it is not easy to disentangle between these two
effects (e.g., Merritt 1987). However, most integrated profiles,
$\sigma_v(<R)$ (i.e., $\sigma_v$ evaluated by using the all the
galaxies within the projected radius $R$), become flat in the external
cluster regions (F96; Girardi et al. 1996).  This suggests that
possible velocity anisotropies does not significantly affect the value
of $\sigma_v$ when it is estimated over a wide cluster region, so as
to be representative of the total kinetic energy.

Figure 3 shows the integrated velocity dispersion
profiles, $\sigma_v(<R)$, for all the CNOC clusters. The clusters
which are sampled up to external cluster regions really shows a
flattening of the profiles (e.g., A2390, MS0451+02, MS1621+26), while
the behavior of profiles for some others is still uncertain (e.g.,
MS1006+12).

%%FIGURE 3%%%
%\end{multicols}
%\begin{figure}
\includegraphics{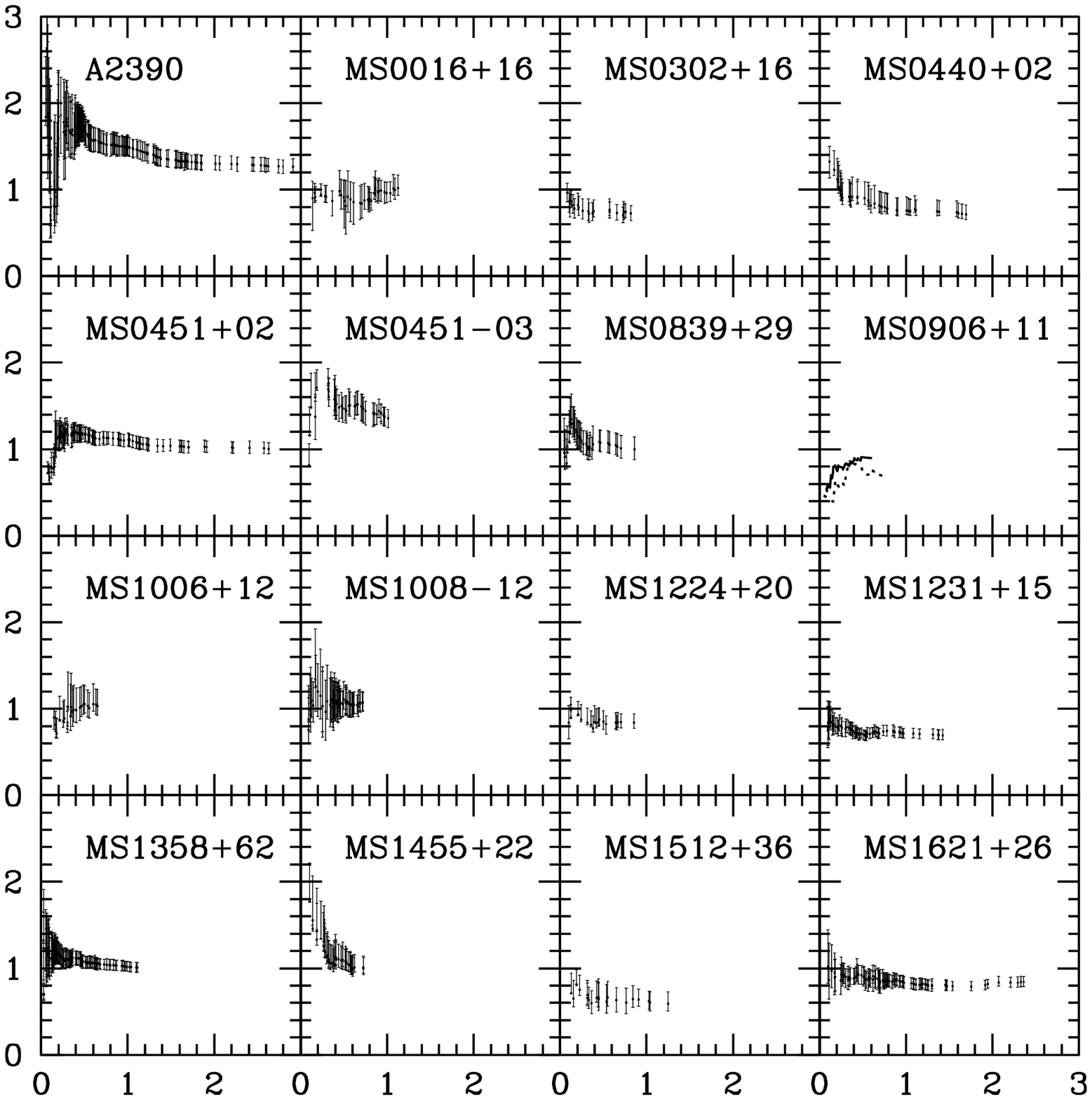}
$\ \ \ \ \ \ $\\
\vspace{9.4truecm}
$\ \ \ $\\
%\vspace{-20mm}
{\small\parindent=3.5mm {Fig.}~3.--- 
Integrated line-of-sight velocity dispersion profiles, where
the dispersion at a given radius is $\sigma_{rob}$ estimated by
considering all galaxies within that radius.  The bootstrap error
bands at the $68\%$ c.l. are shown.  Distances on the $x$--axis are in
units of $\hm$, and the velocity dispersions on the $y$--axis are in
units of $10^3\vel$. As for MS0906+11 the solid and dotted lines
correspond to peak $a$ and $b$, respectively (errorbars are omitted
here for reasons of clarity).
}
\vspace{5mm}
%\begin{multicols}{2}
%%%%%%%%%%%%%%

In order to test the robustness of the $\sigma_v$ computations, we
also use three other estimators: $(a)$ a weighted version of the
robust estimate $\sigma_{w-rob}$, $(b)$ the usual standard deviation
$\sigma_s$, and $(c)$ its weighted version $\sigma_{w-s}$. As for the
weighted estimators, to each galaxy is associated a
magnitude--dependent geometric weight, which has been introduced in
order to account for the geometry of the field surveyed around each
cluster and the magnitude completeness of the sample (cf. Yee,
Ellingson, \& Carlberg 1996 for further details).  The result of the
$\sigma_v$ analysis are reported in Table~1 where we give for each
cluster: the total number of galaxy redshifts, $N_{tot}$, available
for each cluster (Column~2); the number of member galaxies found in
peaks, $N_m$, as recognized by our interloper--removal algorithm
(Column~3); the bi--weight mean cluster velocity (Column~4); the
values of $\sigma_{rob}$ with the relative bootstrap error (at $68\%$
c.l.; Column~5); and the other three estimates of the velocity
dispersion that we considered (Columns~6-8).

%%TAB1

\vspace{6mm}
\hspace{-4mm}
%\begin{minipage}{9cm}
\begin{table*}
\renewcommand{\arraystretch}{1.5}
\renewcommand{\tabcolsep}{2mm}
\begin{center}
\vspace{-3mm}
TABLE 1\\
\vspace{2mm}
{\sc Cluster Velocity Dispersions\\}
\footnotesize
\vspace{2mm}

\begin{tabular}{lccccccc}
\hline \hline
\multicolumn{1}{l}{Name}
&\multicolumn{1}{c}{$N_{tot}$}
&\multicolumn{1}{c}{$N_m$}
&\multicolumn{1}{c}{$V$}
&\multicolumn{1}{c}{$\sigma_{rob}$}
&\multicolumn{1}{c}{$\sigma_{w-rob}$}
&\multicolumn{1}{c}{$\sigma_{s}$}
&\multicolumn{1}{c}{$\sigma_{w-s}$}\\
\multicolumn{1}{c}{}
&\multicolumn{1}{c}{}
&\multicolumn{1}{c}{}
&\multicolumn{1}{c}{$\vel$}
&\multicolumn{1}{c}{$\vel$}
&\multicolumn{1}{c}{$\vel$}
&\multicolumn{1}{c}{$\vel$}
&\multicolumn{1}{c}{$\vel$}
\\
\hline
A2390       &252&191&68608&$1262_{-~68}^{+~89}  $& 1213  &  1306  &1024  \\
MS0016$+$16 &65& 41&164729&$1127_{-112}^{+166}$& 1042  &  1029  &1014  \\
MS0302$+$16 &58& 27&127333&$~710_{-~60}^{+106} $& 688   &  678   &630   \\
MS0440$+$02 &78& 47&58793&$~715_{-~68}^{+113} $& 815   &  709   &542   \\
MS0451$+$02 &189&122&60057&$~1002_{-~61}^{+~72}  $& 1009   &  964   &805   \\
MS0451$-$03 &74& 46&161689&$1330_{-~94}^{+111} $& 1382  &  1281  &1286  \\
MS0839$+$29 &94& 47&57818&$~980_{-113}^{+147}$& 1289  &   964  & 831  \\
MS0906$+$11a&92& 50&52940&$~886_{-~68}^{+~78}  $& 845   &  846   &799   \\
MS0906$+$11b&92& 40&49428&$~725_{-~61}^{+~82}  $& 749   &  691   &697   \\
MS1006$+$12 &38& 30&77965&$1017_{-103}^{+161} $& 1012  &  994  &955  \\
MS1008$-$12 &84& 69&91891&$1042_{-~94}^{+121} $& 1101  &  1060  &1074  \\
MS1224$+$20 &34& 24&97508&$~831_{-~57}^{+129} $& 886   &  777   &779   \\
MS1231$+$15 &120& 77&70329&$~686_{-~50}^{+~65}  $& 660   &  663   &640   \\
MS1358$+$62 &209&136&98404&$1003_{-~52}^{+~61}   $& 1038  &   973  & 917  \\
MS1455$+$22 &68& 51&76991&$1032_{-~95}^{+130} $& 962  &  965  &972  \\
MS1512$+$36 &84& 26&11199&$~575_{-~90}^{+138} $& 717   &  577   &483   \\
MS1621$+$26 &173&101&128182&$~839_{-~53}^{+~67}  $& 811   &  814   &856   \\

\hline
\end{tabular}
 
\end{center}
\vspace{3mm}
%\end{minipage}
\end{table*}
\normalsize

%%FIGURE 4%%%
%\end{multicols}
%\begin{figure}
\vspace{-1.truecm}
\includegraphics{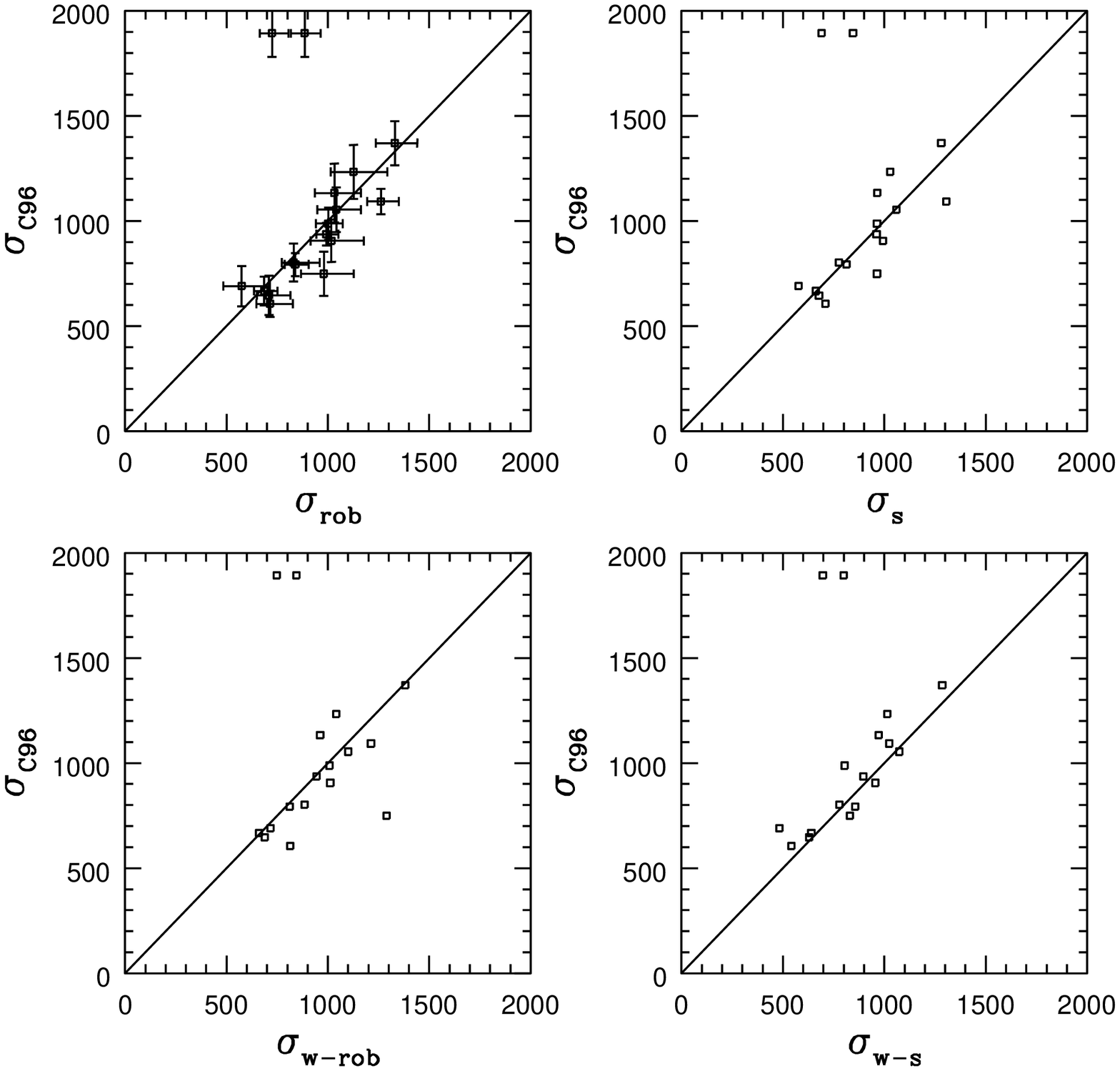}
$\ \ \ \ \ \ $\\
\vspace{9.4truecm}
$\ \ \ $\\
%\vspace{-20mm}
{\small\parindent=3.5mm {Fig.}~4.--- 
The comparison between line-of-sight velocity dispersions (in
units of $\vel$) as estimated by Carlberg et al. (1996),
$\sigma_{C96}$, and the four types of estimates we obtain in this work
(see text). The $1\sigma$ bootstrap errorbars are shown only in the
$\sigma_{C96}$--$\sigma_{rob}$ comparison. The two very discrepant
points refer to the two sub-clusters in which MS0906+11 has been
divided.
}
\vspace{5mm}
%\begin{multicols}{2}
%%%%%%%%%%%%%%

In Figure 4 we compare our estimates of $\sigma_v$ to
those, $\sigma_{C96}$, provided by C96 and based on the explicit
background subtraction algorithm.  Apart from the exception of
MS0906+11, which we divide into two subclumps, the present $\sigma_v$
estimates show an overall agreement with $\sigma_{C96}$. We find that
the median values of the ratio $\sigma_v/\sigma_{C96}$ for the
different estimators lie in the range $0.96-1.04$. In particular, it
is $\sigma_{rob}/\sigma_{C96}= 1.04^{+0.06}_{-0.07}$, where errors are
at $90\%$ c.l.  It is remarkable that different methods give
$\sigma_v$ which, on average, agree within $\mincir 10\%$, thus
confirming the result by Girardi et al. (1993) that different
estimators of $\sigma_v$ give statistically similar results on cluster
samples.  This supports the reliability of such cluster dynamical
studies from optical observations.

Unless otherwise specified, we adopt in the following the
$\sigma_{rob}$ estimator, which is the most directly comparable to
that used by F96 and G98.

We also show in Figure 5 the correlation between
$\sigma_v$ and bolometric $X$--ray luminosity for CNOC clusters
(filled circles). This correlation is also compared to that obtained
for local clusters by combining the XBACs sample (Ebeling et al. 1996)
with the G98 sample of velocity dispersions (open circles).

%%FIGURE 5%%%
%\end{multicols}
%\begin{figure}
\vspace{-0.5truecm}
\includegraphics{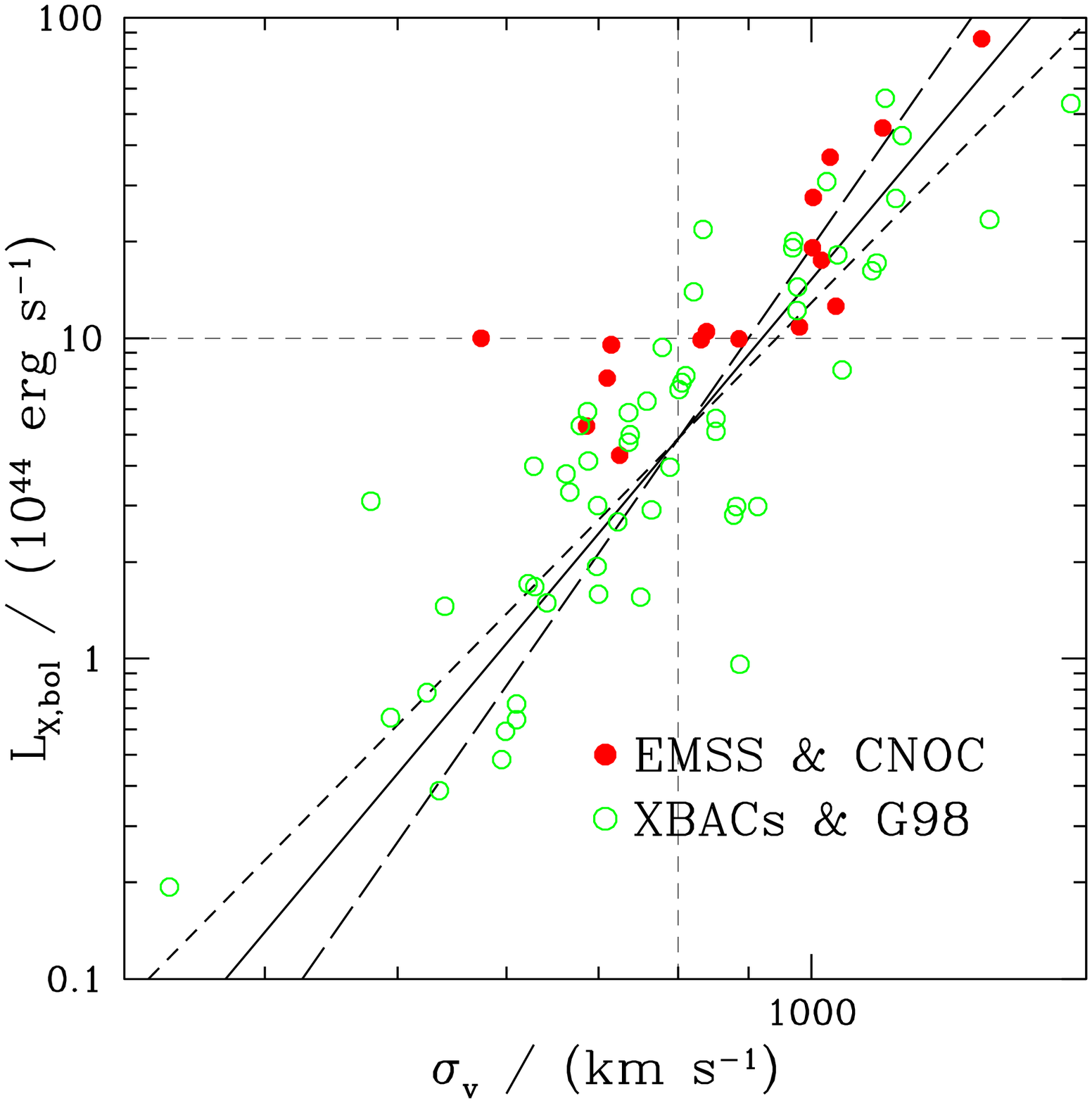}
$\ \ \ \ \ \ $\\
\vspace{9.4truecm}
$\ \ \ $\\
%\vspace{-20mm}
{\small\parindent=3.5mm {Fig.}~5.--- 
The $L_{bol}$--$\sigma_v$ relation for local (open circles)
  and distant CNOC clusters (filled circles). For the local data, we show
  results for those clusters of the Girardi et al. (1998) sample,
  which are also included in XBACs (Ebeling et al. 1996). We
  restricted to those clusters whose $\sigma_v$ is estimated at least
  with 30 galaxy redshifts. For CNOC clusters, $X$--ray
  luminosities have been taken from the EMSS (Henry et al. 1992). 
  Short--dashed, long--dashed
  and continuous lines are the direct, inverse and bisector linear
  regression for the local data points. The light horizontal and
vertical dashed lines indicate the $L_{X,bol}$ and $\sigma_v$ limits
above which we select CNOC clusters for deriving cosmological
constraints.   
}
\vspace{5mm}
%\begin{multicols}{2}
%%%%%%%%%%%%%%

Bolometric luminosities for CNOC (XBACs) clusters, $L_{X,bol}$, have
been estimated by multiplying luminosities in the [0.3--3.5] keV
([0.1--2.4] keV) energy band by a temperature--dependent bolometric
correction factor. This factor has been computed under the assumptions
of pure bremsstrahlung ICM emission and power--law approximation for
the Gaunt factor, $g(E,kT)\propto (E/kT)^{-0.3}$, which is quite
accurate for $kT>2$ keV (cf. B99). Temperatures are available for all
but two CNOC clusters, namely MS1231+15 and MS1621+26 (Lewis et
al. 1999).  For such two clusters they have been estimated from the
velocity dispersion according to $kT=(\sigma_v/350\vel)$ keV. As for
the double cluster MS0906+11, we assigned the same overall
temperature, $kT=8$ keV, to both subclusters. The resulting bolometric
correction factors range from 1.6 to 3.7. According to
Fig. 5, it turns out that there is no appreciable evolution
for the $L_{X,bol}$--$\sigma_v$ relation, thus consistent with the
lack of significant evolution in the $L_{X,bol}$--$T$ relation at
$z\mincir 0.4$ (e.g., Mushotzky \& Scharf 1997) and also in agreement
with the recent analysis by . The only outlier is
MS1512+36, which is one of the two highly substructured objects that
we discussed above. The solid line represents the bisector between the
direct and the inverse log--log linear regression (short-- and
long--dashed lines, respectively) for local clusters. The resulting
$L_X$--$\sigma_v$ relation is given by
\be 
\log L_X\,=\,5.1^{+1.2}_{-0.8}\log \sigma_v -14.2^{+3.0}_{-2.2}
\label{eq:lsig}
\ee 
with a scatter $\Delta_{L_X}=\Delta\log L_X/\log L_X \simeq 0.36$
around the best--fitting relation and in agreement with the recent
analysis by Wu, Xue \& Fang (1999). The upper and lower errors
represent here the difference with respect to the direct and the
inverse linear regression, respectively.  Eq.(\ref{eq:lsig}) has been
obtained by considering the 53 clusters of the G98 sample, which have
$\sigma_v>500\vel$ and at least 30 galaxy members (cf. G98).  If
$T_X\propto \sigma_v^2$, as expected under the assumption of
isothermal gas and hydrostatic equilibrium, then eq.(\ref{eq:lsig})
would imply $L_{X,bol}\propto T_X^{\sim 2.5}$ for the
luminosity--temperature relation, a results which is not far from
recent calibrations of the $L_{X,bol}$--$T_X$ relation for rich
clusters (e.g., Allen \& Fabian 1998; Arnaud \& Evrard 1999, and
references therein).

\section{Constraining $\Omega_m$}

\subsection{Modeling the $\sigma_v$ distribution}
\label{s:ps}
The starting point of this analysis is represented by the
Press--Schechter (1974, PS hereafter) expression for the
comoving number density, at redshift $z$, of virialized halos with mass
in the range $[M,M+dM]$:
\ba
N(M,z)\,dM & = & \sqrt{2\over \pi}\, {\bar \rho \over M^2}\,
{\delta_c(z)\over \sigma_M}\, \left|{d\log \sigma_M\over d\log
M}\right| \nonumber \\
& \times &
\exp\left(-{\delta_c(z)^2\over 2\sigma_M^2}\right)\,dM\,.
\label{eq:ps}
\ea 
Here $\bar \rho$ is the present--day average matter density and
$\delta_c(z)$ is the linear--theory overdensity extrapolated at the
present time for a uniform spherical fluctuation collapsing at
redshift $z$.  It is convenient to express it as
$\delta_c(z)=\delta_{c,0}(z)\,[D(0)/D(z)]$, where $D(z)$ is the linear
fluctuation growth factor (see, e.g., Peebles 1993). For a
critical--density Universe, $\delta_{c,0}=1.686$ with a weak
dependence on $\Omega_m$. We take for $\Omega_m<1$ the expression
provided by Lacey \& Cole (1993) and Kitayama \& Suto (1996).
The quantity
$\sigma_M$ is the present day linear r.m.s. fluctuation within a
sphere of radius $R=(3M/4\pi \bar\rho)^{1/3}$ and is specified by the
power--spectrum, $P(k)$, of density fluctuations. We assume for $P(k)$
the parametric CDM--like expression provided by Bardeen et al. (1986).
Accordingly, its profile and amplitude are determined by the
shape--parameter $\Gamma$ ($\simeq 0.2$ from the the observed galaxy
distribution; e.g., Efstathiou, Bond \& White 1992) and by $\sigma_8$,
the r.m.s. fluctuation amplitude within a sphere of $8\hm$ radius.
The reliability of eq.(\ref{eq:ps}) for predicting the mass distribution
of virialized halos has been tested and debated at length in the
literature (e.g., Lacey \& Cole 1994; Bryan \& Norman 1998; Gross et
al. 1998; B99; Governato et al. 1999) and we refer to such papers for
further details.

In order to convert the distribution of cluster masses into that of
one--dimensional velocity dispersions, $\sigma_v$, we use the relation
\ba
M(\sigma_v)&=&\left({\sigma_v\over 1129\,f_\sigma \vel}\right)^3
\left({\Delta_c(z)\over 178}\right)^{-1/2} \nonumber \\
&\times &E^{-1}(z) \times
10^{15}h^{-1}M_\odot \,,
\label{eq:sigm}
\ea
where $E(z)=[(1+z)^3\Omega_m+(1+z)^2(1-\Omega_m-\Omega_\Lambda)+
\Omega_\Lambda]^{1/2}$ and $\Delta_c$ is the ratio between the average
density within a virialized halo and the critical background density
at $z$ (e.g., Bryan \& Norman 1998). It is $\Delta_c=178$ for
$\Omega_m=1$ while the expression for $\Omega_m<1$ has been taken from
Kitayama \& Suto (1996).
Eq.(\ref{eq:sigm}) corresponds to the case of an isothermal
halo density profile for $f_\sigma=1$ (Binney \& Tremaine 1987), with
slightly smaller values, $0.9\mincir f_\sigma\mincir 1$, for steeper
profiles at the virial radius (e.g., Navarro, Frenk \& White
1996). Since different halos have different dynamical histories, we
expect that an intrinsic scatter, $\Delta_M=\Delta M/M$, should be
present around eq.(\ref{eq:sigm}). Once the $M$--$\sigma_v$ relation
and its scatter are calibrated, the distribution of velocity
dispersions, $N(\sigma_v)$, is obtainable by convolving
eq.(\ref{eq:ps}) with a Gaussian distribution having dispersion
$\Delta_M$. 

In order to estimate $f_{\sigma_v}$ and $\Delta_M$, we resort to
$N$--body simulations, based on the Adaptive P3M code by Couchman
(1991). We analyze two sets of simulations, which have been run for an
Einstein--de-Sitter (EdS) model and for a flat low--density model with
$\Omega_m=0.4$ ($\Lambda$0.4). Both models have been chosen to be
consistent with the abundance of local clusters. Each simulation box
is $250\hm$ aside and five realizations have been run for each
model. Such simulations have already been used by B99 to test the PS
mass function and we refer to that paper for further details.  Once
clusters are identified, $f_\sigma$ and $\Delta_M$ are estimated from
their $M$--$\sigma_v$ virial relation. The results of this analysis
are reported in Figure 6, where we plot, at three
different redshifts, $f_\sigma$ and $\Delta_M$ as a function of the
cluster number density, lower $n_{cl}$ corresponding to the population
of richer clusters.  We find that $f_\sigma$ and $\Delta_M$ are
essentially independent of both the simulated models and the
evolutionary stage. Only the EdS result on $f_\sigma$ at $z=0.2$ for
the smallest $n_{cl}$ (i.e., for the most massive cluster population)
shows a marginal departure from the general behavior. We checked that
this is generated by one single cluster, appearing in one of the five
realizations, which has an anomalous high $\sigma_v$.  Based on this
result, we adopt in the following the best fitting values,
$f_\sigma=0.93$ and $\Delta_M=0.15$.

%%FIGURE 6%%%
%\end{multicols}
%\begin{figure}
\includegraphics{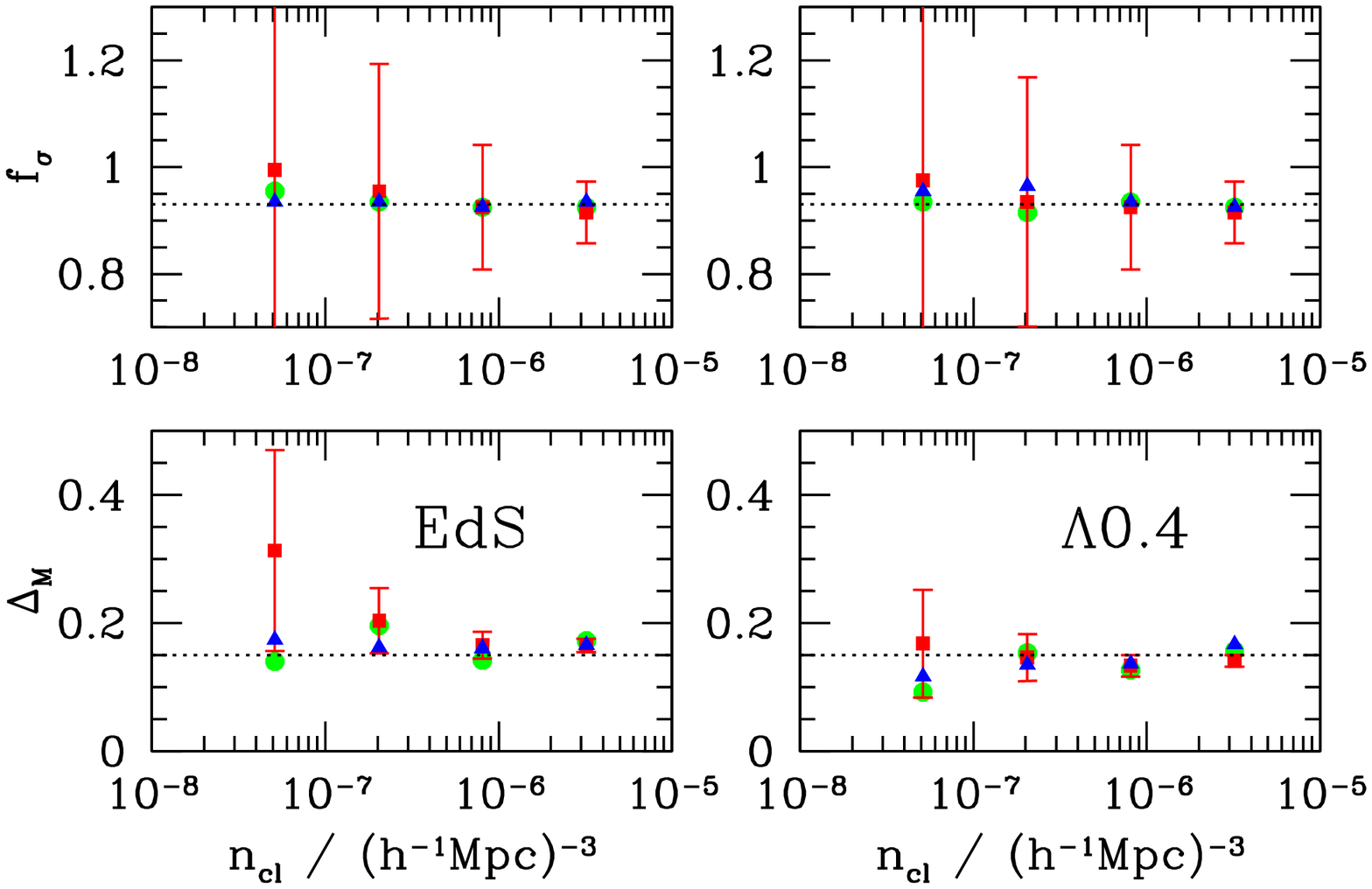}
$\ \ \ \ \ \ $\\
\vspace{-2.5truecm}
\vspace{9.4truecm}
$\ \ \ $\\
%\vspace{-20mm}
{\small\parindent=3.5mm {Fig.}~6.--- The normalization $f_\sigma$
(upper panels) and the scatter $\Delta_M$ (lower panels) for the
$M$--$\sigma_v$ relation [cf.  eq.(\ref{eq:sigm})] from N--body
simulations are plotted against the cluster number density. Left and
right panels are for the EdS and the $\Lambda$0.4 model, respectively
(see text). In each panel, circles, squares and triangles refer to the
outputs at $z=0,0.2,0.6$, respectively. Errorbars, which are reported
only for the $z=0.2$ output, are $1\sigma$ Poisson scatter.  }
\vspace{5mm}
%\begin{multicols}{2}
%%%%%%%%%%%%%%

\subsection{Maximum--likelihood analysis}
\label{s:ml}
In order to constrain cosmological parameters we compare the cluster
distribution on the $(\sigma_v,z)$ plane with model predictions for
samples having the same completeness and selection criteria of the
CNOC sample.  A similar approach has been recently used by Eke et
al. (1998) to follow the evolution of the cluster $X$--ray temperature
function for the Henry (1997) sample, and by Borgani et al. (1998) to
follow the evolution of the $X$--ray luminosity function in the {\sl
ROSAT} Deep Cluster Survey (Rosati et al. 1998).

This method is based on partitioning the $(\sigma_v,z)$ plane into
elements of size $d\sigma_v\,dz$ and on computing the model
probability $\lambda(\sigma_v,z)\,d\sigma_v\,dz$ of observing a CNOC
cluster with velocity dispersion $\sigma_v$ at redshift $z$, given the
completeness and selection criteria of the sample. If the bin width is
small enough that such probabilities are always much smaller than
unity, then the likelihood function ${\cal L}$ of the observed cluster
velocity dispersions and redshift is defined as the product of the
probabilities of observing exactly one cluster in $dz\,d\sigma_v$ at
each of the $(\sigma_{v,i},z_i)$ positions occupied by the CNOC
clusters, and of the probabilities of observing zero clusters in all
the other differential elements of the $(\sigma_v,z)$ plane. Assuming
Poisson statistics for such probabilities, we obtain 
\ba
{\cal L} & = & \prod_{i}\left[\lambda(\sigma_{v,i},z_i,)dz\,d\sigma_v\,
e^{-\lambda(\sigma_{v,i},z_i) dz\,d\sigma_v}\right]\,\nonumber \\
& \times & \prod_{j\ne i}
e^{-\lambda(\sigma_{v,j},z_j)dzd\sigma_v}\,,
\label{eq:like}
\ea
where the indices $i$ and $j$ run over the occupied and empty elements
of the $(\sigma_v,z)$ plane, respectively. Defining, as usual,
$S=-2{\mbox{\rm ln}}{\cal L}$, and dropping all the terms which do not
depend on the parameters of the cosmological model, we find
\ba
S\,=\,-2{\mbox{\rm ln}}{\cal L} & = & -2\sum_i{\mbox{\rm 
ln}}[\lambda(\sigma_{v,i},z_i)]\nonumber \\
& + & 2\int dz \int d\sigma_v\,\lambda(\sigma_v,z)\,,
\label{eq:ent}
\ea 
where the integral is performed over the redshift and velocity
dispersion intervals spanned by the cluster sample (see also Marshall
et al. 1983).

The function $\lambda (\sigma_v,z)$ is given by
\be
\lambda(\sigma_v,z)\,=\,N(\sigma_v,z)\,{dV(z)\over dz}\,f_{sky}\,,
\label{eq:lam}
\ee 
where $N(\sigma_v,z)$ is the comoving number density of clusters
with velocity dispersion $\sigma_v$ at redshift $z$, $dV(z)$ is the
comoving volume element in the redshift interval $[z,z+dz]$ and
$f_{sky}$ is the sky--coverage, i.e. the effective fraction of the sky
surveyed by EMSS.
In order to account for bootstrap errors in the $\sigma_v$ estimates we
compute $N(\sigma_v,z)$ by convolving the Press--Schechter-based
predictions with a 10\% Gaussian--distributed relative scatter, which
is representative of the statistical uncertainties in $\sigma_v$
(cf. Table 1).
Since $f_{sky}$ for EMSS clusters depends on the detect--cell flux, we
compute it for given $\sigma_v$ and $z$ by integrating over all the
possible fluxes, thereby accounting for the scatter in the
$L_{bol}$--$\sigma_v$ correlation of eq.(\ref{eq:lsig}). 
In order to account for the fact that detect--cell fluxes are smaller
than total cluster fluxes, we divide the latter by a factor 2.1, which
has been shown by Henry et al. (1992) to be appropriate at least for
clusters in the redshift range [0.22--0.33] (cf. their Figure 1).

A further selection effect that we have to account for in the estimate
of the likelihood function is connected to the fact that, at a given
$\sigma_v$, the $X$--ray luminosity-- and flux--limit criteria cause
faint clusters to be missed in the CNOC sample.  In order to correct
for this $\sigma_v$--incompleteness, we follow a similar procedure to
that adopted by C97. We select those CNOC clusters which have
$\sigma_v>800\vel$ and $L_{X,bol}>10^{45}\lum$. Then we
resort to a complete local cluster sample to estimate the fraction
${\cal F}_{800}$ of all clusters with
$\sigma_v>800\vel$, which also satisfy the above luminosity
selection.  Finally, we correct for the $\sigma_v$ incompleteness by
multiplying by ${\cal F}_{800}^{-1}$ the sum over the
occupied cells in the r.h.s. of eq.(\ref{eq:ent}).

To estimate ${\cal F}_{800}$, we cross--correlate the ENACS cluster
sample (Katgert et al. 1998), which at $z\mincir 0.1$ is
representative of the whole cluster population for $\sigma_v>800\vel$
(Mazure et al. 1996), with the XBACs (Ebeling et al. 1996), which
includes all the Abell/ACO clusters identified in the {\sl ROSAT}
all--sky survey and is $80\%$ complete for fluxes
$f_{[0.1-2.4]}>5\times 10^{-12}\fl$. Since at $z=0.1$ this flux
corresponds to a luminosity $L_{X,[0.1-2.4]}\simeq 2\times
10^{44}\lum$, we are reasonably guaranteed that, for typical
bolometric corrections, XBACs contains all the Abell--type clusters
with $L_{X,bol}>10^{45}\lum$ in the redshift range covered by
ENACS. Therefore, ${\cal F}_{800}$ is given by the fraction of ENACS
clusters with $\sigma_v>800\vel$ and belonging to the XBACs with
$L_{X,bol}>10^{45}\lum$. It turns out that ENACS contains 26 clusters
with $\sigma_v>800\vel$, out of which 6 have
$L_{X,bol}>10^{45}\lum$. Therefore, we obtain ${\cal
F}_{800}^{-1}=4.3\pm 2.0$, where the uncertainty is the $1\sigma$
Poissonian error. We note that using local data to estimate ${\cal
F}_{800}$ is consistent with the lack of significant
evolution in the $L_{X,bol}$--$\sigma_v$ relation shown in
Fig. 5. 

In their analysis, C97 used the data on $L_X$ and $\sigma_v$ by Edge
\& Stewart (1991) and found ${\cal F}_{800}^{-1}=2.0\pm 1.0$. The
consequence of a smaller ${\cal F}_{800}^{-1}$ is that a smaller
fraction of clusters is expected to be lost below the luminosity
limit. Therefore a smaller number of clusters with $\sigma_v>800\vel$
is expected at high redshift and a larger value of
$\Omega_m$ would be implied. This will be discussed on a more
quantitative ground in \S4.3 below.

Best estimates of the model parameters are obtained by minimizing $S$
and confidence regions are estimated by allowing for standard
increments $\Delta S$. Unless otherwise specified, in the following
we will restrict our analysis only to the case of vanishing
cosmological constant.

\subsection{Results}
\label{s:res}
Figure 7 shows constraints on the $\sigma_8$--$\Omega_m$
plane taking $\Gamma=0.2$ for the power--spectrum shape
(we checked that results are weakly dependent on $\Gamma$ within the
range allowed by galaxy clustering data; e.g., Peacock \& Dodds 1994). 
Such results are based on adopting the
unweighted robust estimator for $\sigma_v$ (cf. Section
\ref{s:data}) and taking only clusters with $\sigma_v>
800\vel$ and $L_{X,bol}>10^{45}\lum$.  The plotted iso--likelihood
contours corresponds to $1\sigma$, $2\sigma$ and $3\sigma$ confidence
levels. The dashed curves are the $\sigma_8$--$\Omega_m$
relation coming from the local cluster abundance and correspond to
$\tilde \sigma_8=0.50$, 0.55 and 0.60, while the shape of the
$\sigma_8$--$\Omega_m$ relation is that provided by Girardi et
al. (1998a).

Both the small number of clusters on which the analysis is based and
the limited leverage in redshift allow a rather high level of
degeneracy in the $\sigma_8$--$\Omega_m$ plane.  As expected, more
stringent constraints on $\Omega_m$ come by combining the results from
CNOC clusters with those from the local cluster abundance.  Taking
$\tilde \sigma_8=0.55$, we find $\Omega_m=0.65_{-0.08}^{+0.13}$ (at
the 2$\sigma$ c.l. for one significant parameter), thus ruling out a
critical density model at a high confidence level (if
$\Omega_\Lambda=1-\Omega_m$, then this result changes into
$\Omega_m=0.53^{+0.16}_{-0.12}$). However, the variation of
the best--fitting $\Omega_m$ as $\tilde \sigma_8$ is increased from
0.5 to 0.6 is much larger than internal statistical uncertainties;
indeed, we find that $\Omega_m= 0.35_{-0.07}^{+0.10}$ and
$1.05_{-0.14}^{+0.20}$ for $\tilde \sigma_8=0.5$ and 0.6,
respectively.  Therefore, although uncertainties from the local
cluster abundance are believed to be rather small, they propagate into
large uncertainties in the determination of $\Omega_m$ from distant
cluster data (cf. also Colafrancesco, Mazzotta \& Vittorio 1997). This
is due to the fact that CNOC clusters are rather rich systems and,
therefore, probe the high--mass tail of the mass function, which is
very sensitive to the power spectrum normalization. On the one hand,
this confirms the good news that results from a small number of
clusters with precise determinations of $\sigma_v$ can be used to
constrain $\Omega_m$, once they are combined with local data. On the
other hand, the bad news is that current uncertainties in the local
cluster abundance are still large enough to prevent placing strong
constraints on the density parameter. We regard this as a general
problem one has to face with, even assuming that the analysis of
high--$z$ sample is perfectly under control.

%%FIGURE 7%%%
%\end{multicols}
%\begin{figure}
\includegraphics{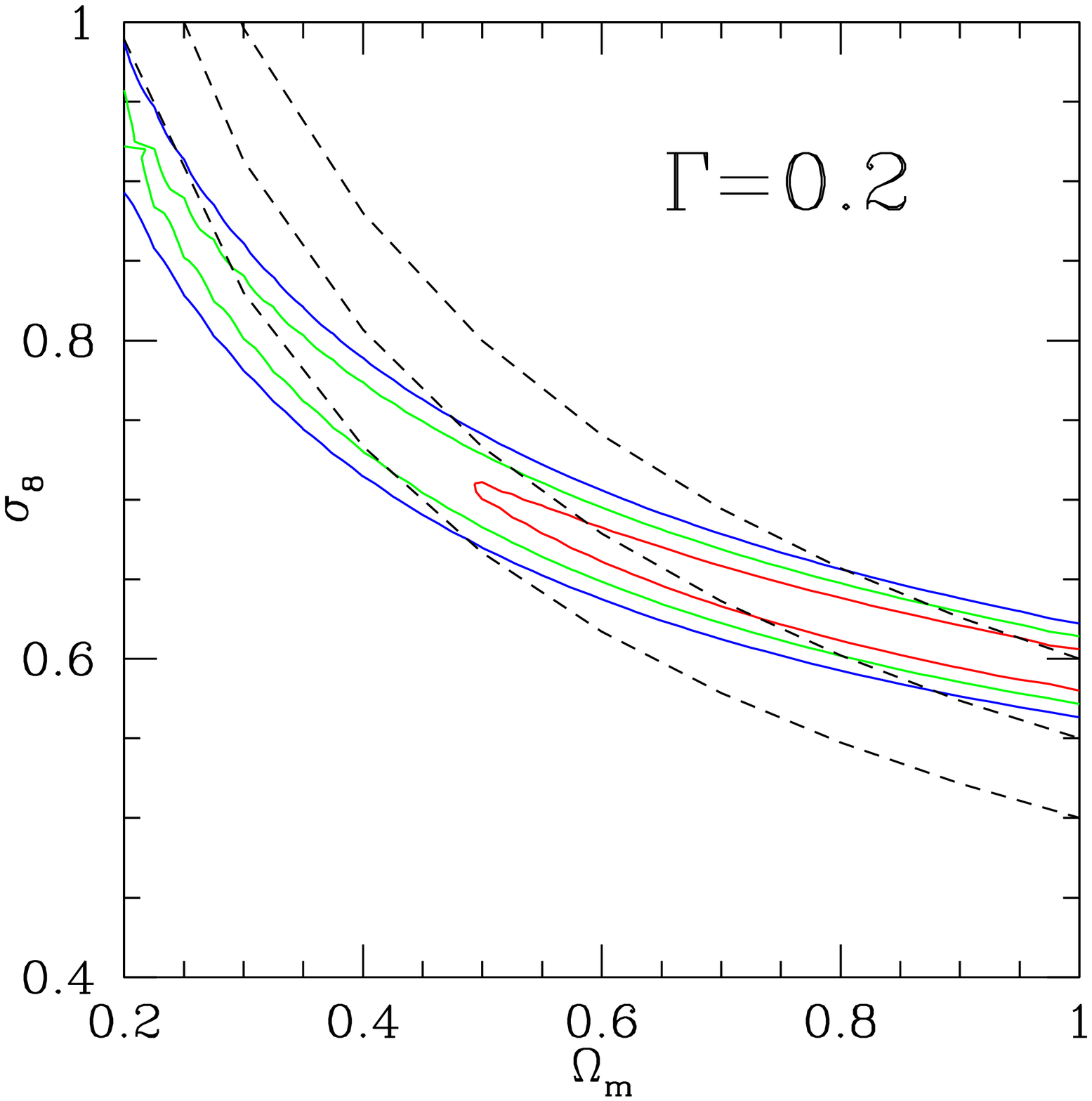}
$\ \ \ \ \ \ $\\
\vspace{9.4truecm}
$\ \ \ $\\
%\vspace{-20mm}
{\small\parindent=3.5mm {Fig.}~7.--- 
Iso--likelihood contours on the $(\sigma_8,\Omega_m)$ plane
for $\Gamma=0.2$. Results refer to the reference analysis method, as
described in Section \ref{s:res}. Iso--likelihood contours are for
$\Delta S=2.30$,6.17 and 11.8, which corresponds to 1$\sigma$,
$2\sigma$ and $3\sigma$ c.l. for two significant fitting
parameters. The three dashed curves are the $\sigma_8$--$\Omega_m$
relation from the local cluster abundance, for
$\tilde\sigma_8=0.50,0.55$ and 0.60, from lower to upper curves, while
the scaling is taken from Girardi et al. (1998b).
}
\vspace{5mm}
%\begin{multicols}{2}
%\end{figure}
%%%%%%%%%%%%%%

A further source of uncertainty in constraining $\Omega_m$ arises from
the correction for the $\sigma_v$--incompleteness. We find that, as
${\cal F}_{800}$ varies, within its $1\sigma$ range, from 2.3 to 6.3
(cf. \S 4.2), the best--fitting $\Omega_m$ decreases from $\simeq 0.9$
to $\simeq 0.5$.  Therefore, the effect of this uncertainty is also
non--negligible and calls for the need of a substantially larger
number of clusters with both $L_X$ and $\sigma_v$ determinations to
suppress the Poissonian uncertainty in the estimate of ${\cal F}_{800}$.

Furthermore, we note from Fig. 5 that there are two
clusters, MS1224+20 and MS0906+11a, whose values of $L_{X,bol}$ are
only slightly below the adopted luminosity limit. Owing to residual
uncertainties in the bolometric correction, one may wander by how much
results would change if such clusters were included in the
analysis. In this case, we find that, for $\tilde\sigma_8=0.55$, the
best--fitting $\Omega_m$ only decreases by 0.06. 

In order to clarify the crucial role of the low--$z$ normalization for
tracing the redshift evolution of the cluster abundance, we compare in
Figure 8 the comoving cluster number density, with
$\sigma_v>800\vel$, to model predictions for a critical and an open
low--density model, assuming different normalizations from the local
cluster abundance. The estimate of the CNOC cluster abundance is obtained
by dividing the nine selected clusters into two redshift intervals, $0.15
\le z \le 0.30$ and $0.30\le z\le 0.55$, which contain five and four
clusters, respectively. Number densities are then estimated by
applying the $1/V_{max}$ method (Avni \& Bahcall 1978) and following
the same procedure as C97. Apart from the details of the model
constraints, which are in general agreement with the results from
Fig. 7, this plot gives a visual impression of why a
change in the normalization from the local cluster abundance turns
into a change of the preferred $\Omega_m$ value. For instance,
$n(>800,z=0)\simeq 3\times 10^{-6}(\hm)^{-3}$, as coming from the
virial analysis of local clusters (e.g., Mazure et al. 1996; Fadda et
al. 1996), would be consistent with a critical density Universe, while
$n(>800,z=0)\simeq 10^{-6}(\hm)^{-3}$, as implied by the estimates of
the local $X$--ray temperature function by Eke et al. (1996) and
Markevitch (1998), would instead favor a low--density Universe.

%%FIGURE 8%%%
%\end{multicols}
\begin{figure*}[t]
\vspace{-1.5truecm}
\includegraphics{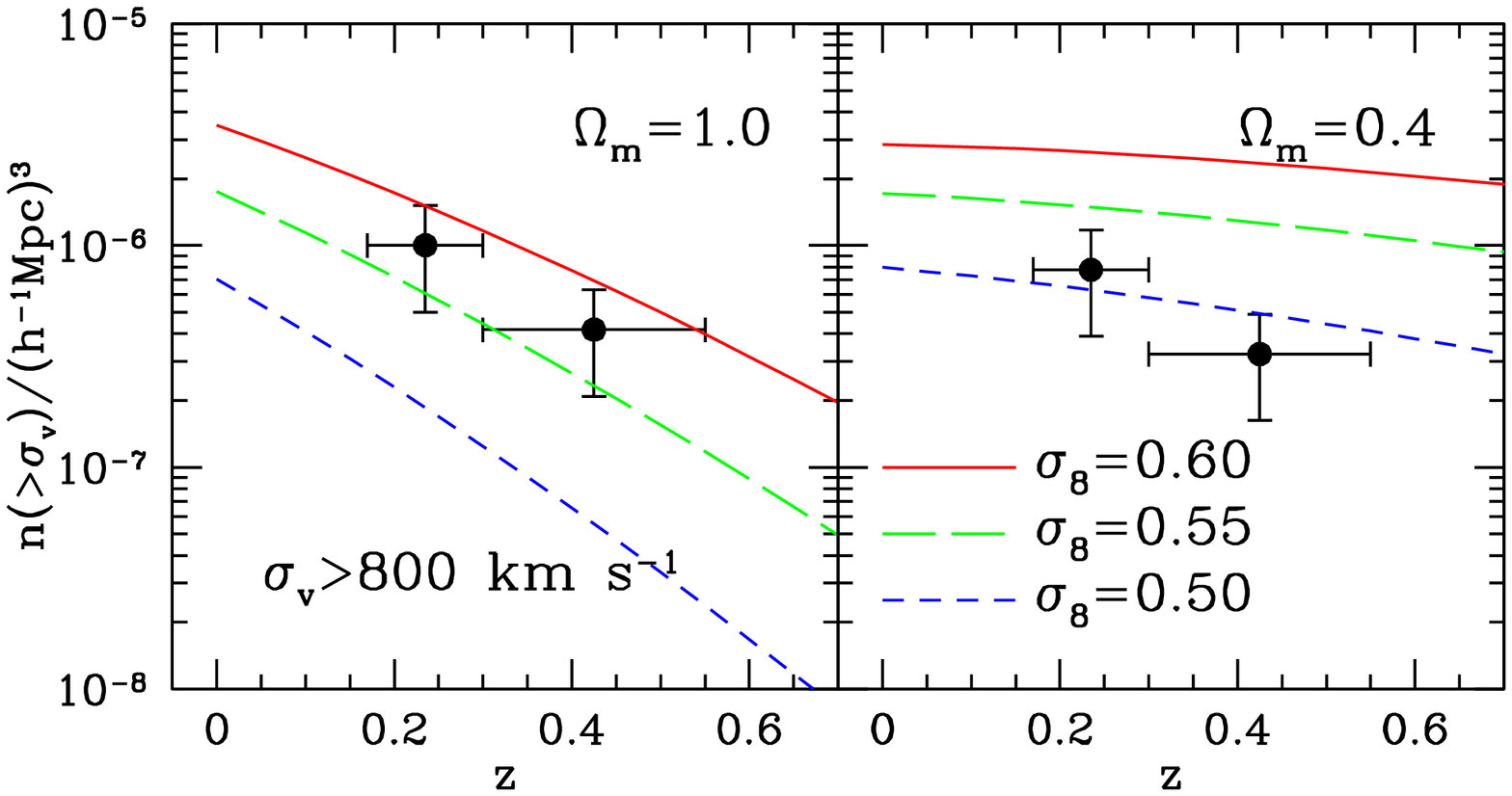}
$\ \ \ \ \ \ $\\
\vspace{9.4truecm}
$\ \ \ $\\
{\small\parindent=3.5mm {Fig.}~8.--- 
The redshift evolution of the comoving number
density of CNOC clusters with $\sigma_v>800\vel$, compared with model
predictions.  Left and right panels refer to a critical--density and
to an open model with $\Omega_m=0.4$, respectively. In each panel, the
three different curves refer to three different normalizations from
the local cluster abundance. We assume $\Gamma=0.2$ for the shape
parameter of the power spectrum. The filled dots are the CNOC cluster
number density in two redshift bins (see text). Vertical errorbars
are $1\sigma$ Poissonian uncertainties in $n(>\sigma_v)$, while
horizontal errorbars indicate the extent of the redshift bins.
}
\vspace{5mm}
%\begin{multicols}{2}
\end{figure*}
%%%%%%%%%%%%%%

Since this is not the first analysis aimed at constraining $\Omega_m$
and $\sigma_8$ from the CNOC cluster abundance evolution (C97; Bahcall
et al. 1997), it is worth comparing our results to those from such
previous analyses. C97 provided the number density of clusters with
mass within a physical radius of $1.5\hm$ above a given limit by
dividing the CNOC sample into two redshift bins. It is interesting to
note that, despite the value of ${\cal F}_{800}$ used in that analysis
is about twice as large than ours, their results prefer lower
$\Omega_m$ (cf. Bahcall et al. 1998, for a stronger claim for a low
density parameter, $\Omega_m\simeq 0.3$, from CNOC clusters). The
reason for this difference is mainly due to the fact that C97 use a
fixed normalization for the local cluster abundance. In particular,
they noticed that, if ENACS velocity dispersions are thought to be
overestimated by 13\%, then the implied cluster abundance is
compatible with that found by Eke et al. (1996) from the $X$--ray
temperature function. In fact, the number density of local clusters
with $\sigma_v>800\vel$, as given by C97, would be compatible with
$\tilde\sigma_8=0.5$. In view of the good agreement we found here
between velocity dispersions estimated with the explicit background
subtraction and with the method applied to ENACS clusters, we argue that
there is no evidence that ENACS velocity dispersions should be 
overestimated.

\section{Conclusions}
We analyzed the internal velocity dispersions, $\sigma_v$, of CNOC
clusters (e.g., Yee, Ellingson \& Carlberg 1996), by applying the
algorithm originally developed by Fadda et al. (1996, F96; cf. also
Girardi et al. 1998b, G98) for the analysis of local clusters. After
removing interlopers, we applied four different $\sigma_v$
estimators. By using the robust estimator, we found that for
$\sigma_v> 800\vel$ and $L_{X,bol}>10^{45}\lum$ the CNOC sample
contains nine clusters, on which we base our analysis. In order to
account for the clusters that we miss with respect to an ideal sample
complete for $\sigma_v>800\vel$, we resort to local data in order to
reliably estimate the fraction ${\cal F}_{800}$ of clusters that
in such a sample have $L_{X,bol}>10^{45}\lum$.
The resulting $\sigma_v$ are used to trace the evolution of
the cluster abundance in the $0.17\mincir z \mincir 0.55$ redshift
range probed by the CNOC sample, with the purpose of placing
constraints on the $\sigma_8$--$\Omega_m$ plane. Such constraints can
then be combined with results from the local cluster abundance to
constrain the cosmological density parameter. 

The main results of our analysis can be summarized as follows.

\begin{description}
\item[(a)] The explicit background subtraction method applied by
Carlberg et al. (1996, C96) on CNOC clusters provides $\sigma_v$
estimates which are fully consistent with those provided by the method
applied in this paper. For instance, we find that the median of the
ratio between the robust estimator and the C96 results is
$\sigma_{rob}/\sigma_{C96}=1.04^{+0.06}_{-0.07}$.
\item[(b)] Current uncertainties in the
$\sigma_8$--$\Omega_m$ relation from the local cluster abundance are
still large enough not to allow CNOC data to distinguish to a high
confidence level among a low--density Universe with $\Omega_m\simeq
0.3$ and a critical density scenario. 
\end{description}

Far from meaning that the evolution of the cluster abundance can
hardly be used to place significant constraints on cosmological
parameters, the results of our analysis anyway indicate that the
situation is more complex than sometimes suggested. Somewhat
surprisingly, the availability of precise cluster mass measurements at
high redshift through both $\sigma_v$ and $X$--ray temperature
measurements may be only of partial help in increasing the strength of the
constraints unless {\em (a)} systematic uncertainties in the local
mass--function are reduced at a $\mincir 10\%$ level, and {\em (b)}
the completeness criteria of high--$z$ cluster samples are well under
control.

\acknowledgments We acknowledge stimulating discussions with Neta
Bahcall, Andrea Biviano, Alain Blanchard, Isabella Gioia, Michael
Gross, Simon Morris and Joel Primack. We thank Hugh Couchman for the
sharing of his Adaptive P3M N--body code. SB and MG acknowledge SISSA
in Trieste for its hospitality during the preparation of this work.

%%%%%%%%%%%% per formato preprint
\end{multicols}
%%%%%%%%%%%% per formato preprint   

%\begin{multicols}{2}

\end{document}